\documentclass[aps
,preprint
,groupedaddress
]{revtex4-1}
\usepackage{graphicx}     
\usepackage{fancyhdr}
\usepackage[colorlinks,linkcolor=black,anchorcolor=blue,citecolor=red]{hyperref}
\usepackage{amsthm}
\usepackage{amsmath}      

\DeclareMathOperator*{\diag}{diag}


\newtheorem{thm:def}{Definition}[section]
\newtheorem{thm:thm}{Theorem}[section]
\newtheorem{thm:lemma}{Lemma}[section]
\newtheorem{thm:rmk}{Remark}[section]
\newtheorem{thm:corollary}{Corollary}[section]

\begin{document}
\title{Flow Decomposition Reveals Dynamical Structure of Markov Process}
\author{Jianghong Shi}
\author{Tianqi Chen}
\author{Bo Yuan}
\email[Corresponding author. Email: ]{boyuan@sjtu.edu.cn}
\affiliation{Department of Computer Science and Engineering, Shanghai Jiao Tong University, Shanghai, 200240, China.}
\author{Ping Ao}
\email[Corresponding author. Email: ]{aoping@sjtu.edu.cn}
\affiliation{Shanghai Center for Systems Biomedicine, Key Laboratory of Systems Biomedicine of Ministry of Education, Shanghai Jiao Tong University, Shanghai, 200240, China.\\
Institute of Theoretical Physics, Shanghai Jiao Tong University, Shanghai, 200240, China.}

\date{\today}

\begin{abstract}
Markov process is widely applied in almost all aspects of literature, especially important for understanding non-equilibrium processes. We introduce a decomposition to general Markov process in this paper. This decomposition decomposes the process into 3 independent parts: stationary distribution, symmetric detailed-balance part and anti-symmetric breaking detailed-balance part. This complete decomposition captures the steady state as well as the dynamics of the process, providing an elegant perspective for construction or analyzing problems. In light of the decomposition, a unique definition of relative entropy is found to formally separate the effect of detailed-balance part and breaking detailed-balance part. We find that the relative Gini entropy production introduced in the paper is not affected by the non-detailed balance part of the process. This property do not holds for other entropy definition in general discrete case.
\end{abstract}

\maketitle

\section{Introduction}
Markov process is widely applied in almost all aspects of sciences, especially in biology,
from complex networks \cite{qian_2010,wang_2008, ao_cancer_2010,ao_cancer_2008} to
evolution dynamics \cite{jiao_2011,xu_2011,zhou_2011}, and many others \cite{sleep_2009,squartini_2008,cell_2011,ao_phage_2004}.

In general, we can group Markov process into two categories, detailed balance Markov process
and non-detailed balance Markov process.  The detailed balance Markov process is well studied and applied in literature. On the other hand, most systems found in nature are not detailed balance ones. Such kind of systems include biological systems, chemical systems, economical systems etc. Due to the lack of detailed balance condition, the studies for these systems are much harder than detailed balance ones.

For general Markov process, it is a good way to decompose it into separate parts for categorizing and analyzing purpose. Many attempts have been taken to decompose Markov process to reveal its underlying structure \cite{sch_1976}\cite{qian_1979}\cite{jiang_2004}\cite{zia_2007}. In this paper, we introduce a flow decomposition view of general Markov process, whose special case in continuous space embraces a recent developed framework in terms of Stochastic differential equation\cite{ao_2005,ao_2006,ao_2007,ao_2008}.

Our decomposition decomposes the process into 3 independent parts: stationary distribution, symmetric detailed-balance part and anti-symmetric breaking detailed balance part (circulation flow). This complete decomposition captures the steady state as well as the dynamics of the process, providing an elegant perspective for construction or analyzing problems. In this paper, we concentrate on discussing the properties of a class of state functions under the decomposition.

State functions are widely used when considering macroscopic properties of complex systems. It is well known that Shannon entropy does not monotonically increase in general Markov process \cite{van_2007}. Instead, relative entropy or free energy preserve this monotone property. In addition, \cite{ao_2008} demonstrates that in continuous space Markov process, the antisymmetric part does not affect the derivative of all definitions of relative entropy. In discrete space, however, this claim is no longer valid for all definitions of relative entropy. Among them, we find a definition of relative entropy (we name it relative Gini entropy) to formally separate the effect of detailed-balance part and breaking detailed balance part in general Markov process, in both discrete and continuous spaces. The specialty of relative Gini entropy lies in that it is the only definition of relative entropy in discrete space that the effect of breaking detailed balance part would vanish. Definitions such as relative Shannon entropy do not enjoy this property and will lead to subtle difficulties \cite{ge_2009} when analyzing. To have a taste of the power of relative Gini entropy, we demonstrate that the relative Gini entropy production bound can leads to famous eigen value bound \cite{diaconis_1996}.

This article is structured as follows. In the second section, we will first
review the concept of Markov process we discuss in this article. Then we will talk about
our flow decomposition of Markov process in the third section. The fourth section serves
as an example to use the framework to analyze the entropy evolution in non-detailed balance process. At last, we will talk about connections of our work to others and give conclusion and discussions.

\section{Flow Decomposition on Master Equation}
In this section, we introduction our flow decomposition method on master equation.
We will limit our discussion to continuous time Markov process and restrict our process to be with a unique stationary distribution which is non-zero over all the points in the state pace.

\subsection{Master equation}
Assume the discrete state space $S=\{1,2,\cdots n\}$. The discrete
space Markov process is usually described as
following master equation \cite{sch_1976}.
\begin{equation}
  \label{eq:cme_sum}
  \partial_t p_i(t) = \sum_{j\neq i} \mathbf{Q}_{ij} p_j(t) + \mathbf{Q}_{ii} p_i(t),
  \mathbf{Q}_{ii} = - \sum_{j\neq i} \mathbf{Q}_{ji}
\end{equation}
Where $\mathbf{Q_{ij}}\geq 0(i\neq j)$ are transition rate constants.
The master equation have natural interpretation as follows: the
change of probability in a state is equal to the inflow of
probability from other states minus the outflow from current state
to others. Here $\partial_t p_i(t)$ corresponds to the change of
probability in state $i$. $\mathbf{Q}_{ij}p_j(t)$ the inflow from
$j$ to $i$, and $-\mathbf{Q}_{ii}p_i(t)$ is the total outflow from
state $i$. Equation \ref{eq:cme_sum} is usually described in matrix
form
\begin{equation}
  \label{eq:cme_mat}
  \partial_t p(t) = \mathbf{Q} p(t)
\end{equation}
The stationary distribution can be solved by setting
\begin{equation}
  \label{eq:cme_stat}
  0 = \mathbf{Q} \pi
\end{equation}
Because $\mathbf{Q}$ is not a full rank matrix, solution of Equation \ref{eq:cme_stat}
must exist and is eigen-vector of $\mathbf{Q}$. The solution is unique when $\mathbf{Q}$
only have one zero eigen-value. There is an elegant graph algorithm to calculate the stationary distribution \cite{sch_1976}\cite{zia_2007}. Other more intuitive conditions on $\mathbf{Q}$ also exists to guarantee the uniqueness of the solution. As stated before, we will assume such conditions exist for our problem.
For futher details of the conditions,
readers can refer to \cite{gardiner}\cite{van_2007}.

\subsection{Flow decomposition on master equation}
The flow decomposition of master equation \ref{eq:cme_mat} is formally given by the following equations.
\begin{equation}\begin{split}\label{eq:flow_decomp_discrete}
  \partial_t p(t) &= \mathbf{Q} \;p(t) \\
                  &= \mathbf{F} \; \mathbf{\Pi}^{-1} \; p(t) \\
                  &= [\mathbf{S}+\mathbf{A}] \; \mathbf{\Pi}^{-1} \; p(t), \\
   where \; \mathbf{\Pi} & = \diag[\pi_1, \pi_2,\dots, \pi_n], \\
    \mathbf{S}   &= (\mathbf{F}+\mathbf{F}^{\tau})/2, \; \mathbf{A}= (\mathbf{F}-\mathbf{F}^{\tau})/2
\end{split}\end{equation}
Here $\mathbf{\Pi}$ is a diagonal matrix with the stationary distribution $\pi$ as its elements. $\mathbf{F}_{ij} =\mathbf{Q}_{ij} \pi_j $ is the probability flow
from state $j$ to state $i$ in the stationary distribution, thus has zero column sum and row sum. $\mathbf{S}$ is a symmetric matrix and $\mathbf{A}$ is an anti-symmetric matrix, both of which have zero column sum and row sum.

Each part of the decomposition have their own meanings. With the probability conservation nature of master equation, we just decompose the probability flow $\mathbf{F}$ into two different parts. The symmetric detailed balance part $\mathbf{S}$ has an equal
probability flow between arbitrary pair of states. While the
anti-symmetric breaking detailed balance part $\mathbf{A}$ stands for circulation flux.

In cases when the circulation flux is vanishing, i.e.,
$\mathbf{A}=0$, we can find the process satisfies the detailed
balance condition. That is why we call $\mathbf{S}$ detailed balance
part and $\mathbf{A}$ breaking detailed balance part. Noting that the
degree of freedom of $\mathbf{A}$ is zero when $n=2$, which means
that a system of 2 states always meets detailed balance condition.
It is only when $n \geq 3$ that the circulation flux emerges.

With simple calculation, it can be verified that the total degree of
freedom of the three parts is same as original $\mathbf{Q}$ matrix.
In fact, the symmetric matrix $\mathbf{S}$ with $n(n-1)/2$ degree of
freedom, the anti-symmetric matrix $\mathbf{A}$ with $(n-1)(n-2)/2$
degree of freedom, and the stationary distribution $\pi$ with $n-1$
degree of freedom which are mutually independent add up to the
degree of freedom of Q matrix $n^2-n$.

\begin{figure}
\centering
\includegraphics[scale = 0.5]{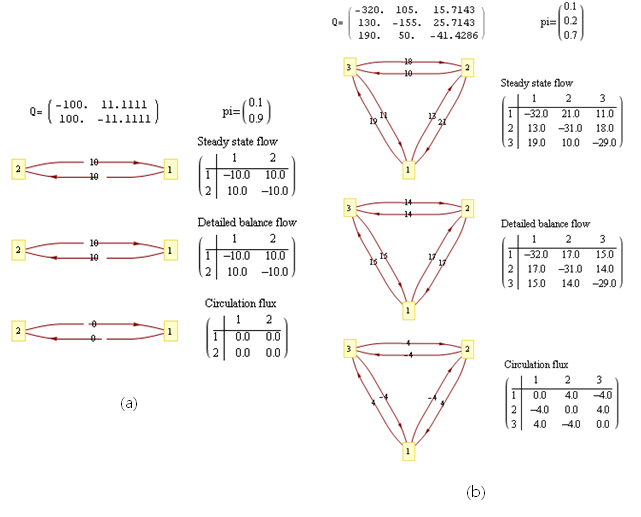}
\caption{\textbf{Flow decomposition.} \textbf{a.} 2 states
decomposition. \textbf{b.} 3 states
decomposition.}\label{fig:decomposition}
\end{figure}

Figure \ref{fig:decomposition} shows flow decomposition with 2-state system and 3-state system. If there are only 2 states, the total degree of freedom of Q-matrix is 2, which include
1 for stationary distribution, and 1 for the symmetric flow, noting
that in this case anti-symmetric flux vanishes. In the 3-state case,
the anti-symmetric flux emerges with 1 degree of freedom. The
detailed balance flow has 3 degree of freedom and stationary
distribution has 2. They add up to 6 degrees of freedom of the
original Q-matrix.

A straightforward advantage of the decomposition view is that it
reveals more detailed information by each part. For example, we can
easily construct a series of Markov process with same stationary
distribution $\pi$ by changing $\mathbf{S}$ and $\mathbf{A}$. It
also gives some connection between the process and its dual process. We
can get the dual process by reversing anti-symmetric flow
$\mathbf{A}$.

\section{Abstract Flow Decomposition}
Now we introduce an abstract flow decomposition framework here, which includes both discrete and continuous space Markov processes as special cases.

 A general Markov process can be formalized as following equation
\begin{equation}
  \partial_t \rho(t) =  \mathcal{L} [\frac{\rho(t)}{\pi} ]
\end{equation}
Where $\mathcal{L}\frac{1}{\pi}$ is called generator. $\rho$ is the
probability density function in continuous space case and probability
function in discrete space case. Abstract flow decomposition says
that we can decompose operator $\mathcal{L}$ into two operators.
\begin{equation}
  \mathcal{L} = \mathcal{S} + \mathcal{A}
\end{equation}

$\mathcal{S}$ is an self-adjoint operator and $\mathcal{A}$ is
anti-symmetric operator. They satisfies the following properties:
\begin{equation}
  \langle \mathcal{S}u, v \rangle =   \langle u ,\mathcal{S} v \rangle
\end{equation}
\begin{equation}
  \langle \mathcal{A}u, v \rangle =   -\langle u ,\mathcal{A} v \rangle
\end{equation}
The inner product is Euclid inner product in the space of $\rho$.
The flow decomposition in the discrete case is a special case of
this abstract definition. Where $\mathbf{Q}\pi$ corresponds to the
operator $\mathcal{L}$, and $\mathbf{A}$ and $\mathbf{S}$
corresponds to $\mathcal{A}$ and $\mathcal{S}$.

This abstract definition serves as generalization of the finite
discrete space decomposition. While we can still view $\mathcal{S}$,
$\mathcal{A}$ and $\pi$ as detailed balance flow, non-detailed
balance flow and stationary distribution. We show that this
abstract decomposition includes the previous framework of
decomposition in continuous case in appendix.

\section{Dynamics in Terms of Entropy Evolution}
Entropy is one of the most important and mysterious state function
in thermodynamics. Discussions on Entropy has been going for
hundreds of years and are still going on. How to define entropy in
non-equilibrium systems is still one of the biggest problems.

It is well known that Shannon entropy in Markov process is not monotonically increasing. While a general definition of relative entropy makes second law of thermodynamics a built-in property of master equation described system. Previous work \cite{ao_2008} has shown that in continuous space case, the derivative of any definition of relative entropy is not affected by the breaking detailed part. Nevertheless, this is no longer true in the more general discrete space case, where only one special definition of relative entropy (relative Gini entropy) enjoys this property. We will discuss relative Gini entropy and show its power by using it to prove the famous eigen value bound.

\subsection{Entropy and relative entropy}
The most famous definition of entropy is the Shannon entropy introduced
from information theory and statistical physics \cite{cover_1991}.
\begin{equation}
  H(p) = -\sum_{i=1}^n p_i \ln p_i
\end{equation}
However, this definition of entropy is commonly not monotone in general Markov process. Accordingly, applying  maximum entropy principle with this definition of entropy to general process probably will end up with inconsistent conclusions. On the other hand, if we take the stationary distribution into account and introduce the definition of relative entropy, the monotone property will be preserved and the second law of thermodynamics(arrow of time) will turn out to be a built-in property for master equation described processes.

For a discrete state space with $n$ state, the relative $f$ entropy is defined as follows
\begin{equation}
  S^{f}(p) = -\sum_{i=1}^n \pi_i f \left(\frac{p_i}{\pi_i}\right)
\end{equation}
Here $f$ is a convex function. Different choices of $f$ give
different definitions of entropy. The most common one is relative
Shannon entropy.

The monotone property of relative entropy is well known in literature, the following Figure.\ref{fig:entropy_mono} serves as an example to illustrate the underlying reason of how this happens. (a)Compares the evolution of the non-monotone Shannon entropy(Dashed lines) and the monotone Relative Shannon entropy(solid lined). Different colors are for different
initial distribution. (b)and(c) further illustrates how (a) happens: The non-uniform stationary distribution cause a shifting of the fix state from the center, leading to the sometimes increasing and sometimes decreasing Shannon entropy.
\begin{figure}
\centering
\includegraphics[scale = 0.5]{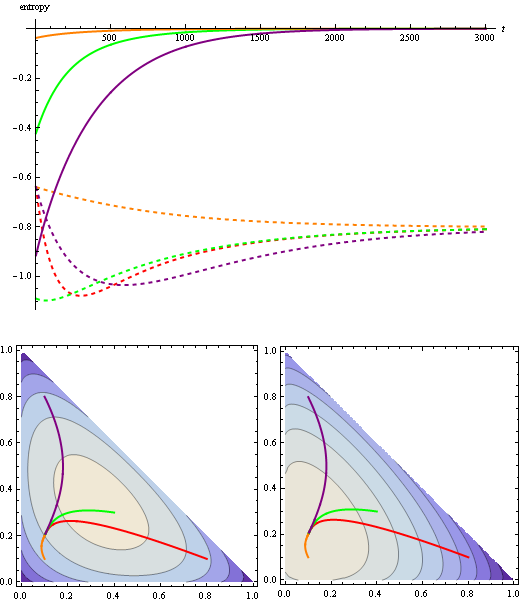}
\caption{\textbf{Entropy evolution.} \textbf{a.} Monotonic evolution
of Relative Shannon Entropy(Solid) and Non-monotonic evolution of
Shannon Entropy(Dashed).Different colors are for different initial
distributions. \textbf{b.} System evolution in Shannon entropy
contours. \textbf{c.} System evolution in Relative Shannon entropy
contours. } \label{fig:entropy_mono}
\end{figure}

\subsection{Relative Gini Entropy}
Now we introduce a special definition of relative entropy, the relative Gini entropy. This definition formally separates the effects of detailed balance part and breaking detailed balance part in current decomposition view, suggesting their dissipative and conserved nature separately. It is worth to mention that Gini entropy is the unique choice to have this property in both discrete and continuous space cases. In light of relative Gini entropy, we rewrite the proof of famous eigen value bound to have a taste of the importance of this definition.

Relative Gini entropy \cite{rao_1982}, which is a special case of Tsallis entropy \cite{beck_2009}, chooses  $f(x)=x^2$ and is defined as follows
\begin{equation}
  S^{g}(p) = - \sum_{i=1}^N \pi_i \left(\frac{p_i}{\pi_i}\right)^2
\end{equation}

Using our definition of flow decomposition, we can put the
entropy production of relative Gini entropy as follows
\begin{equation}\begin{split}
  \partial_t S^{g}(p) &= - 2\left(\frac{p}{\pi}\right)^T (\mathbf{S}+\mathbf{A})\left(\frac{p}{\pi}\right) \\
 & = - 2\left(\frac{p}{\pi}\right)^T \mathbf{S}\left(\frac{p}{\pi}\right)- 2\left(\frac{p}{\pi}\right)^T \mathbf{A}\left(\frac{p}{\pi}\right)
\end{split}\end{equation}
Here $\frac{p}{\pi} =
[\frac{p_1}{\pi_1},\frac{p_2}{\pi_2},\cdots,\frac{p_n}{\pi_n}]^T $. Remember that $\mathbf{S}$ is symmetric negative semi-definite matrix and $\mathbf{A}$ is anti-symmetric matrix and they both have zero column sum and role sum. We have
\begin{equation}\label{eq:epr}
  \partial_t S^{g}(p) = - 2\left(\frac{p}{\pi}\right)^T \mathbf{S}\left(\frac{p}{\pi}\right) \geq 0
\end{equation}
Equation \ref{eq:epr} says that relative Gini entropy will monotonically
increase over time, this is exactly the statement of second law of
thermodynamics. In addition, the derivative of relative Gini entropy
production is not affected by the breaking detailed balance part of the
process. This property indicates the conservative nature of the dynamics associated with the breaking detailed balance part. It is worth mentioning that this specialty of relative Gini entropy only exists in discrete case. When it comes to continuous case, other definition of relative entropy can also eliminate the effects of breaking detailed balance part \cite{ao_2008}.

In the following, we give out the proof of eigen value bound in terms of decomposition and relative Gini entropy view.
\begin{thm:thm}
The evolution of relative Gini entropy can be bounded by
\begin{equation}\label{eq:ep_bound}
  S^{g}(p,t) \geq S^{g}(p,0) e^{-\lambda_2 t}
\end{equation}
Where $\lambda_2$ is the second smallest absolute eigen value of the following matrix $\mathbf{G}$
\begin{equation}
\mathbf{G} = \diag(\sqrt \pi)^{-1} \; \mathbf{S} \; \diag(\sqrt \pi)^{-1}
\end{equation}
\end{thm:thm}

\begin{proof}
We know that evolution of the entropy satisfies the following equation.
\begin{equation}
  S^{g}(p,t) = S^{g}(p,0) + \int_{0}^t \partial_t S^{g}(p) dt
\end{equation}
If we can bound the entropy production $\partial_t S^{g}(p)$, we can
bound the rate of convergence of Markov process.

Since we have
\begin{align}
  \partial_t S^{g}(p)
  &= - \left(\frac{p}{\pi}\right)^T \mathbf{S}\left(\frac{p}{\pi}\right)\\
  &= - \left(\frac{p}{\pi}-e\right)^T \mathbf{S}\left(\frac{p}{\pi}-e\right) \\
  &= - \left(\frac{p}{\sqrt{\pi}}-\sqrt{\pi}\right)^T \mathbf{G} \left(\frac{p}{\sqrt{\pi}}-\sqrt{\pi}\right)
\end{align}
We can find the length(norm) of the vector $\left(\frac{p}{\sqrt{\pi}}-\sqrt{\pi}\right)$ is related to current
 relative Gini entropy
\begin{equation}
  \left|\left|\left(\frac{p}{\sqrt{\pi}}-\sqrt{\pi}\right)\right|\right| ^2
  = \sum_{i=1}^N \frac{p_i^2}{\pi_i} - 2 \sum_{i=1}^N p_{i} + \sum_{i=1}^N \pi_i = - S^{g}(p)
\end{equation}
For the new matrix $\mathbf{G}$, we can find it is also negative semi-definite, and have a eigen-value of 0.
\begin{equation}
  \mathbf{G}\sqrt{\pi} = 0
\end{equation}
Assume the eigen decomposition of $\mathbf{G}$ as follows

\begin{equation}
  \mathbf{G} = - \mathbf{U} \diag([\lambda_1,\lambda_2,\cdots,\lambda_N]) \mathbf{U}^T,
\ \lambda_N \geq \cdots \geq \lambda_2 \geq \lambda_1 = 0, \mathbf{U}_1 = \sqrt{\pi}
\end{equation}

Then the entropy production becomes
\begin{align}
  \partial_t S^{g}(p)
&= \left(\frac{p}{\sqrt{\pi}}-\sqrt{\pi}\right)^T \mathbf{U} \diag([\lambda_1,\lambda_2,\cdots,\lambda_N]) \mathbf{U}^T \left(\frac{p}{\sqrt{\pi}}-\sqrt{\pi}\right)\\
&= y^T \diag([\lambda_1,\lambda_2,\cdots,\lambda_N]) y\\
&= \sum_{i=1}^N \lambda_i y_i^2
\end{align}

Where $y_i$ is the projection of $\left(\frac{p}{\sqrt{\pi}}-\sqrt{\pi}\right)$ over $\mathbf{U}_i$, we can find $y$
have following properties

\begin{equation}
  \left|\left|y\right|\right| ^2
  =\left|\left|\left(\frac{p}{\sqrt{\pi}}-\sqrt{\pi}\right)\right|\right| ^2
  = - S^{g}(p)
\end{equation}
\begin{equation}
  y_1 = \mathbf{U_1}^T \left(\frac{p}{\sqrt{\pi}}-\sqrt{\pi}\right) = \sqrt{\pi}^T \left(\frac{p}{\sqrt{\pi}}-\sqrt{\pi}\right) = 0
\end{equation}

Combining above results, we can state the eigen-value bound of entropy production as Equation.\ref{eq:ep_bound}.
\end{proof}

Due to the nice property of relative Gini entropy, we derive the
well known eigen-value bound of convergence without any struggle. We
find that the previous proofs\cite{fill_1991}\cite{diaconis_1996} of this result implicitly applied the idea of decomposition. This is usually given by the concept of dual
process. Previous study on non-detailed balance process
mostly reduce the problem to the detailed balance part of the
process. Example of such works include eigen-value bound\cite{fill_1991} and
log-Soblev bound\cite{diaconis_1996} of Markov process. Because anti-symmetric part is
not explicitly used, the decomposition can be used implicitly by
concept of process and its dual process.

\section{Discussion}
The abstract flow decomposition we introduced here can be viewed as a generalization of previous work done on continuous space case\cite{ao_2008}\cite{ao_2007}. People
once doubted whether similar method can be applied to discrete space case as well. We at this point illustrated that discrete space case and continuous space case can be unified with abstract flow decomposition view. Since continuous case can be seen as a limit of general discrete case, the conclusions we obtain in discrete case should always be valid in continuous case, but not vice versa. Relative Gini entropy serves as an example here, which is the unique choice to eliminate the effects of breaking detailed balance part in discrete case. While in continuous case, all kinds of relative $f$ entropy can do. Another important thing is that in
discrete case the stationary distribution is not ambiguous, while in continuous case, the emergence
of multiplicative noise will lead to ambiguous stationary distributions \cite{shi_2011}.

The decomposition view of Markov process can act as guide lines for designing algorithms or analyzing problems. Though finding rules in general process is usually difficult, there are already some special cases that can be understood clearly under this view \cite{todo_2010}\cite{diaconis_2000}. More rules are waiting to be discovered along the line. It is worthy to mention that the breaking detailed balance part can be further decomposed into smaller circulation parts \cite{jiang_2004,kalpazidou_1995}, which may be useful for more detailed analysis.

There has been similar work done on decomposition of Markov process in literature \cite{zia_2007}. However, they focus on the steady states. Since our decomposition here is complete, the stationary distribution, the symmetric detailed balance part, and the anti-symmetric breaking detailed balance part together make the whole process, including the information of both steady state and dynamics.

The entropy evolution is also the main concern in literature \cite{ge_2009}\cite{jiang_2004}. For processes with non-uniform distribution as stationary distribution, one major understanding is to decompose the change of the non-monotone Shannon entropy into two parts: non-negative entropy production within the system and transfer of heat across the boundary. This ``outsider'' view is in contrast with our ``insider'' view, which directly consider the relative entropy within the system. Such direct treatment should be easier in dealing with some complicated problems.
Furthermore, relative Gini entropy play a quite special role in light of current decomposition view, which separate the effects of detailed balance part and breaking detailed balance part, leading to simplification for analysis.

\section{Conclusion}
In this work, we introduce a flow decomposition view in dealing with Markov process. It combines the existing decomposition in continuous Markov process and the current decomposition in discrete case. This decomposition view captures both steady states and dynamics structure of Markov process. In light of the decomposition, we find a special definition of relative entropy which can formally eliminate the effects of breaking detailed balance part and lead to dramatic simplification of analysis. We suggest this view as a good perspective for dealing with Markov process, further study should be carried on along the line.

\begin{acknowledgments}
The authors would like to express their sincere gratitude for the helpful discussions with Ruoshi Yuan, Song Xu, Xinan Wang, Yian Ma, Ying Tang.
This work was supported in part by the National 973 Projects No.~2007CB914700 and No.~2010CB529200 (P.A.); and by the Chinese Natural Science Foundation No.~NFSC61073087 (R.Y., J.S. and B.Y.).
\end{acknowledgments}

\appendix
\section{Flow decomposition in continuous process}
The decomposition on continuous space case is put forward by Ao \cite{ao_2005} with the following special form of Fokker-Planck equation, which is the continuous counterpart of master equation.
\begin{equation}
  \label{eq:fpe_decomp}
  \partial_t \rho(x,t) = \nabla^T \left\{\left[  \mathbf{D(x)}+\mathbf{G(x)} \right]\left[ \nabla \phi(x) + \nabla \right]\rho(x,t)\right\}
\end{equation}
Where $x$ is a $n$ dimensional variable, $\mathbf{D(x)}$ is the symmetric positive semi-definite diffusion matrix, $\mathbf{G(x)}$ is anti-symmetric matrix and $\phi(x)$ is the potential function. It is easy to see the stationary distribution is the famous Boltzmann-Gibbs distribution
\begin{equation}
  \pi(x) \propto e^{- \phi(x)}
\end{equation}

The abstract flow decomposition of Equation \ref{eq:fpe_decomp} is given as follows
\begin{equation}
  \mathcal{S}(u) = \nabla^T \left\{\mathbf{D}\left[ \nabla \phi + \nabla \right]
    \pi u\right\}
\end{equation}
\begin{equation}
  \mathcal{A}(u) = \nabla^T \left\{\mathbf{G}\left[ \nabla \phi + \nabla \right]
    \pi u\right\}
\end{equation}
We now prove that operator $\mathcal{S}$ and $\mathcal{A}$ do satisfies
the property of symmetry and anti-symmetry in our definition.

\begin{proof}
  The general idea of proof
\begin{itemize}
\item
  \begin{equation}
    \partial^i\pi(x) = -\pi(x) \partial^i\phi(x)
  \end{equation}
\item Maybe we need to swap entropy $i$ and $j$
\item Use integral by parts when necessary
\end{itemize}
Here $\partial^i$ mean taking partial derivative to $i-th$ dimension.
\begin{align}
  &\langle \mathcal{S}f, g \rangle\\
 =& \int g(x) \sum_i \partial^i \left[\sum_j D_{ij} f(x)\pi(x) \partial^j\phi(x) + \sum_j D_{ij} f(x) \partial^j\left(f(x)\pi(x)\right)\right] dx\\
 =& \sum_{ij}\int g(x) \partial^i \left[\textcolor{red}{D_{ij} f(x)\pi(x) \partial^j\phi(x)} + D_{ij}  \partial^j\left(f(x)\pi(x)\right)\right] dx\\
 =& \sum_{ij}\int g(x) \partial^i \left[\textcolor{red}{-D_{ij} f(x) \partial^j\pi(x)} + \textcolor{blue}{D_{ij}  \partial^j\left(f(x)\pi(x)\right)}\right] dx\\
 =& \sum_{ij}\int g(x) \partial^i \left[D_{ij}\pi(x) \partial^j f(x)\right] dx \ \  \mbox{combine red and blue}\\
 =&- \sum_{ij} \int  D_{ij}\pi(x) \partial^j f(x) \partial^i g(x) dx  \ \ \mbox{integral by parts}\\
 =&- \sum_{ji} \int  D_{ji}\pi(x) \partial^i f(x) \partial^j g(x) dx  \ \ \mbox{swap index}\\
 =&\langle f, \mathcal{S} g\rangle  \ \ \mbox{by symmetry}
\end{align}
For $\mathcal{A}$, the proof is exactly the same except using the fact that $G$ is antisymmetric matrix
in swap index step.
\end{proof}

By far we have shown that our definition of flow decomposition
connects to the previous works of decomposition in continuous
diffusion process. This also gives another justification of the flow
decomposition view since we can have a framework working with both
discrete and continuous state space. And our definition of
decomposition in finite discrete Markov process can be viewed as a
discrete space counterpart of the continuous decomposition by
\cite{ao_2008}.

\bibliographystyle{plain}
\bibliography{bib_decomposition}

\end{document}